\documentclass[%
final%
,a4paper%
,mathtime,mtbold,mtpluscal%
]{aipproc}
\layoutstyle{8x11single}
\graphicspath{{\jobname/}}

\newcommand{\nd}          {\discretionary{--}{--}{--}}
\newcommand{\dS}          {\mbox{de Sitter}}
\newcommand{\adS}         {\mbox{anti-de Sitter}}
\newcommand{\Sch}         {Schwarz\-schild}
\newcommand{\SdS}         {\Sch\nd\dS}
\newcommand{\SadS}        {\Sch\nd\adS}

\newcommand{\RN}          {Reissner\nd Nordstr\"om}
\newcommand{\RNdS}        {\RN\nd\dS}

\newcommand{\KN}          {Kerr\nd Newman}
\newcommand{\CLT}         {Coriolis\nd Lense\nd Thirring}
\newcommand{\ed}          {embedding diagram} 
\newcommand{\eds}         {\ed s}
\newcommand{\dif}         {\mathrm{d}}
\newcommand{\oder}[2]     {\frac{\dif #1}{\dif #2}}
\newcommand{\loder}[2]    {\dif #1/\dif #2}

\newcommand{\eulern}      {\mathrm{e}}
\newcommand{\eto}[1]      {\eulern^{#1}}
\newcommand{\text}[1]     {\mathrm{#1}}

\newcommand{\tfrac}[2] {{\textstyle\frac{#1}{#2}}}
\newcommand{\takenfrom}[1]{Taken from \citep{#1}.}
\renewcommand{\vec}[1]    {\mbox{\boldmath$#1$}}
\let\idxI=\kappa \let\idxII=\lambda \let\idxIII=\mu
\begin{document}

\title{Visualizing spacetimes via embedding diagrams}

              

\author{Stanislav Hled\'{\i}k}{
  address={Institute of Physics, Faculty of Philosophy \& Science, Silesian
  University in Opava,\\ CZ-746\,01 Opava, Czech Republic}
}
\author{Zden\v{e}k Stuchl\'{\i}k}{
  address={Institute of Physics, Faculty of Philosophy \& Science, Silesian
  University in Opava,\\ CZ-746\,01 Opava, Czech Republic}
}
\author{Alois Cipko}{
  address={Institute of Physics, Faculty of Philosophy \& Science, Silesian
  University in Opava,\\ CZ-746\,01 Opava, Czech Republic}
}

\begin{abstract}
  It is hard to imagine curved spacetimes of General Relativity. A~simple but
  powerful way how to achieve this is visualizing them via \emph{embedding
    diagrams} of both ordinary geometry and optical reference geometry.  They
  facilitate to gain an intuitive insight into the gravitational field
  rendered into a curved spacetime, and to assess the influence of parameters
  like electric charge and spin of a black hole, magnetic field or
  cosmological constant. Optical reference geometry and related inertial
  forces and their relationship to embedding diagrams are particularly useful
  for investigation of test particles motion.  Embedding diagrams of static
  and spherically symmetric, or stationary and axially symmetric black-hole
  and \mbox{naked-singularity} spacetimes thus present a useful concept for
  intuitive understanding of these spacetimes' nature. We concentrate on
  general way of embedding into 3-dimensional Euclidean space, and give a set
  of illustrative examples.
\end{abstract}

\maketitle

\section{Introduction}\label{intro}%%%%%%%%%%%%%%%%%%%%%%%%%%%%%%%%%%%%%%%%

The analysis of embedding diagrams
\citep{Stu-Hle:1999:PHYSR4:,Stu-etal:2001:PHYSR4:,Kri-Son-Abr:1998:GENRG2:,Stu-Hle:2001:FPFSU:PropRNdS,Stu-Hle:1999:CLAQG:,Bar:1973:BlaHol:,Stu:2001:RAGtime2and3:,Sla:2001:RAGtime2and3:}
rank among the most fundamental techniques that enable understanding phenomena
present in extremely strong gravitational fields of black holes and other
compact objects.  The influence of charge, spin, magnetic field or
cosmological constant on the structure of spacetimes can suitably be
demonstrated by embedding diagrams of 2-dimensional sections of the ordinary
geometry ($t = \mathrm{const}$ hypersurfaces) into \mbox{3-dimensional}
Euclidean geometry and their relation to the well known `\Sch{} throat'
\citep{Mis-Tho-Whe:1973:Gra:}.

Properties of the motion of both massive and massless test particles can be
properly understood in the framework of optical reference geometry allowing
introduction of the concept of gravitational and inertial forces in the
framework of general relativity in a natural way
\citep{Abr-Car-Las:1988:GENRG2:,Abr-Nur-Wex:1995:CLAQG:,Abr:1990:MONNR:,Abr:1992:MONNR:,Abr-Bic:1991:GENRG2:,Abr-Mil:1990:MONNR:,Mil:1993:RenGRCos:,Abr-Pra:1990:MONNR:,Abr-Nur-Wex:1993:CLAQG:},
providing a description of relativistic dynamics in accord with Newtonian
intuition.

The optical geometry results from an appropriate conformal ($3+1$) splitting,
reflecting certain hidden properties of the spacetimes under consideration
through their geodesic structure.  Let us recall the basic properties of the
optical geometry. The geodesics of the optical geometry related to static
spacetimes coincide with trajectories of light, thus being `optically
straight' \citep{Abr-Pra:1990:MONNR:,Abr-Mil-Stu:1993:PHYSR4:}.  Moreover, the
geodesics are `dynamically straight,' because test particles moving along them
are held by a velocity-independent force \citep{Abr:1990:MONNR:}; they are
also `inertially straight,' because gyroscopes carried along them do not
precess along the direction of motion \citep{Abr:1992:MONNR:}.

Some fundamental properties of the optical geometry can be appropriately
demonstrated namely by embedding diagrams of its representative sections
\citep{Abr-Car-Las:1988:GENRG2:,Kri-Son-Abr:1998:GENRG2:,Stu-Hle:1999:CLAQG:}.
Because we are familiar with the Euclidean space, 2-dimensional sections of
the optical space are usually embedded into the 3-dimensional Euclidean space.
In spherically symmetric static spacetimes, the central planes are the most
convenient for embedding --~with no loss of generality one can consider the
equatorial plane, choosing the coordinate system such that $\theta = \pi/2$.
In \KN{} backgrounds, the most representative section is the equatorial plane,
which is the symmetry plane. This plane is also of great astrophysical
importance, particularly in connection with the theory of accretion disks
\citep{Bar:1973:BlaHol:}.

In spherically symmetric spacetimes (\Sch{} \citep{Abr-Pra:1990:MONNR:},
\SdS{} \citep{Stu-Hle:1999:PHYSR4:}, \RN{} \citep{Kri-Son-Abr:1998:GENRG2:},
and \RNdS{} \citep{Stu-Hle:2001:FPFSU:PropRNdS}), an interesting coincidence
appears: the turning points of the \mbox{central-plane} embedding diagrams of
the optical space and the photon circular orbits are located exactly at the
radii where the centrifugal force, related to the optical space, vanishes and
reverses sign. The same conclusion holds for the Ernst spacetime
\citep{Stu-Hle:1999:CLAQG:}.

However, in the rotating black-hole backgrounds, the centrifugal force does
not vanish at the radii of photon circular orbits
\citep{Iye-Pra:1993:CLAQG:,Stu-Hle-Jur:2000:CLAQG:}.  Of course, the same
statement is true if these rotating backgrounds carry a non-zero electric
charge.

Throughout the paper, geometrical units ($c = G = 1$) are used.

\section{Optical geometry and inertial forces}\label{ogif}%%%%%%%%%%%%%%%%%

The notions of the optical reference geometry and the related inertial forces
are convenient for spacetimes with symmetries, particularly for stationary
(static) and axisymmetric (spherically symmetric) ones. However, they can be
introduced for a general spacetime lacking any symmetry
\citep{Abr-Nur-Wex:1993:CLAQG:}.

Assuming a hypersurface globally orthogonal to a timelike unit vector field
$\vec{n}$ and a scalar field $\Phi$ satisfying the conditions $n_{[\idxI}
\nabla\!_\idxII n_{\idxIII]} = 0$, $n^\idxI n_\idxI = -1$, $\dot{n}_\idxII
\equiv n^\idxI \nabla\!_\idxI n_\idxII = \nabla\!_\idxII \Phi$, the 4-velocity
$\vec{u}$ of a test particle of rest mass $m$ can be uniquely decomposed as
\begin{equation}                                                     
  u^\idxI = \gamma (n^\idxI + v\tau^\idxI)\,.              \label{tpa013e7}
\end{equation}
Here, $\vec{\tau}$ is a unit vector orthogonal to $\vec{n}$, $v$ is the speed,
and $\gamma = (1-v^2)^{-1/2}$.

Introducing, based on the approach of \citet{Abr-Nur-Wex:1993:CLAQG:}, a
projected \mbox{3-space} orthogonal to $\vec{n}$ with the positive definite
metric $h_{\idxI\idxII}$ giving the so-called ordinary projected geometry, and 
the optical geometry $\tilde{h}_{\idxI\idxII}$ by conformal rescaling
\begin{equation}                                          
  \tilde{h}_{\idxI\idxII} =
    \eto{-2\Phi} h_{\idxI\idxII}\,,\qquad
  h_{\idxI\idxII} = g_{\idxI\idxII} + n_\idxI  n_\idxII\,, \label{tpa013e9}
\end{equation}
the projection of the 4-acceleration $a^\bot_\idxI = h^\idxII_\idxI\,
u^\idxIII \nabla\!_\idxIII u_\idxII$ can be uniquely decomposed into terms
proportional to the zeroth, first, and second powers of $v$, respectively, and
the velocity change $\dot{v} = (\eto{\Phi}\gamma v)_{,\idxIII}\,u^\idxIII$.
Thus, we arrive at a covariant definition of inertial forces analogous to the
Newtonian physics \citep{Abr-Nur-Wex:1993:CLAQG:,Agu-etal:1996:CLAQG:RaN}
\begin{equation}
  ma^\bot_\idxI =
    G_\idxI(v^0) +
    C_\idxI(v^1) +
    Z_\idxI(v^2) +
    E_\idxI(\dot{v})\,,
\end{equation}
where the terms
%% \begin{eqnarray}  
%%   G_\idxI&=&-m\nabla\!_\idxI \Phi = - m \Phi\!_{,\idxI}\,, \label{tpa013gf}\\
%%   C_\idxI &=& - m \gamma^2 v n^\idxII
%%     (\nabla\!_\idxII \tau_\idxI -
%%     \nabla\!_\idxI \tau_\idxII)\,,                       \label{tpa013cltf}\\
%%   Z_\idxI &=&
%%     -m (\gamma v)^2
%%     \tilde{\tau}^\idxII
%%     \tilde{\nabla}\!_\idxII \tilde{\tau}_\idxI\,,          \label{tpa013cf}\\
%%   E_\idxI &=& - m \dot{v} \tilde{\tau}_\idxI\,,            \label{tpa013ef}
%% \end{eqnarray}
\begin{equation}
  G_\idxI=-m\nabla\!_\idxI \Phi = - m \Phi\!_{,\idxI}\,,\quad
  C_\idxI = - m \gamma^2 v n^\idxII
    (\nabla\!_\idxII \tau_\idxI -
    \nabla\!_\idxI \tau_\idxII)\,,\quad
  Z_\idxI =
    -m (\gamma v)^2
    \tilde{\tau}^\idxII
    \tilde{\nabla}\!_\idxII \tilde{\tau}_\idxI\,,\quad
  E_\idxI = - m \dot{v} \tilde{\tau}_\idxI\,,                \label{tpa013}
\end{equation}
correspond to the gravitational, \CLT, centrifugal and Euler force,
respectively. Here, $\tilde{\vec{\tau}}$ is the unit vector along $\vec{\tau}$
in the optical geometry and $\tilde{\vec{\nabla}}$ is the covariant derivative
with respect to the optical geometry.

In the simple case of static spacetimes with a field of timelike Killing
vectors $\vec{n}\equiv\vec{\xi}_{(t)} = \partial/\partial t$, we are dealing
with space components that we will denote by Latin indices in the following.
The metric coefficients of the optical reference geometry are given by the
formula
\begin{equation}
  \tilde{h}_{ik} = \eulern^{-2\Phi} h_{ik}\,,\qquad
  \eulern^{2\Phi} = -g_{\mu\nu} \xi^{\mu}_{(t)}\xi^{\nu}_{(t)}
            = - g_{tt}\,.                                   \label{RNdSe73}
\end{equation}

In the optical geometry, we can define a 3-momentum of a test particle and a
3-force acting on the particle \citep{Abr-Nur-Wex:1993:CLAQG:,Stu:1990:BULAI:}
\begin{equation}
  \tilde{p}^i = \eulern^{2\Phi} p^i\,,\qquad
  \tilde{f}_i = \eulern^{2\Phi} f_i\,,                      \label{RNdSe79}
\end{equation}
where $p^i$, $f_i$ are space components of the 4-momentum $\vec{p}$ and the
4-force $\vec{f}$.  Instead of the full spacetime form, the equation of motion
takes the following form in the optical geometry,
\begin{equation}
  m\tilde{f}_i = \tilde{p}^k \tilde{\nabla}\!_k \tilde{p}_i
           + \tfrac{1}{2} m^2\tilde{\nabla}\!_i \Phi\,,  \label{RNdSe80}
\end{equation}
where $\tilde{\nabla}\!_k$ represents the covariant derivative components with
respect to the optical geometry. We can see directly that photon trajectories
($m=0$) are geodesics of the optical geometry. The first term on the
right-hand side of (\ref{RNdSe80}) corresponds to the centrifugal force, the
second one corresponds to the gravitational force.

In the case of Kerr--Newman spacetimes, detailed analysis (see, e.g.,
\citep{Stu-Hle-Jur:2000:CLAQG:}) leads to decomposition into gravitational,
\CLT{} and centrifugal forces (\ref{tpa013})
\begin{eqnarray}                                               
  G_\idxI &=&
    -m \Phi\!_{,\idxI} =
    -m\frac{1}{2}\partial_\idxI
    \left[\ln\left(\frac{g^2_{t\phi} -
      g_{tt}g_{\phi\phi}}{g_{\phi\phi}}\right)
    \right]\,,\\
  C_\idxI %&=&
    &=&mA^2 (\Omega - \Omega_\mathrm{LNRF}) \sqrt{g_{\phi\phi}}
    \left[
      \partial_\idxI\left(g_{t\phi}g_{\phi\phi}^{-1/2}\right) +
      \Omega_\mathrm{LNRF}\,\partial_\idxI
      \sqrt{g_{\phi\phi}}                                 \label{tpa013e22}
    \right]\,,\\
  Z_\idxI %&=&
    &=&-\frac{1}{2} mA^2
    (\Omega - \Omega_\mathrm{LNRF})^2
    g_{\phi\phi}\,
    \partial_\idxI
    \left[
      \ln\left(
           \frac{g^2_{\phi\phi}}{g^2_{t\phi}-g_{tt}g_{\phi\phi}}
         \right)
    \right]\,,                                            \label{tpa013e23}
\end{eqnarray}
respectively; we also denote $\vec{\iota} = \eto{-\Phi}\vec{n}$.  The Euler
force will appear for $\Omega\neq \mathrm{const}$ only, being determined by
$\dot{\Omega} = u^\idxII \nabla\!_\idxII \Omega$ (see
\citep{Abr-Nur-Wex:1995:CLAQG:}). Here we shall concentrate on the inertial
forces acting on the motion in the equatorial plane ($\theta = \pi/2$) and
their relation to the embedding diagram of the equatorial plane of the optical
geometry.

Clearly, by definition, the gravitational force is independent of the orbiting
particle velocity. On the other hand, both the \CLT{} and the centrifugal
force vanish (at any $r$) if $\Omega = \Omega_\mathrm{LNRF}$, i.e., if the
orbiting particle is stationary at the LNRF located at the radius of the
circular orbit; this fact clearly illustrates that the LNRF are properly
chosen for the definition of the optical geometry and inertial forces in
accord with Newtonian intuition.  Moreover, both forces vanish at some radii
independently of $\Omega$. In the case of the centrifugal force, namely this
property will be imprinted into the structure of the embedding diagrams.

\section{Embedding diagrams}\label{ed}%%%%%%%%%%%%%%%%%%%%%%%%%%%%%%%%%%%%%

The properties of the (optical reference) geometry can conveniently be
represented by embedding of the equatorial (symmetry) plane into the
3-dimensional Euclidean space with line element expressed in the cylindrical
coordinates ($\rho,z,\alpha$) in the standard form.  The embedding diagram is
characterized by the embedding formula $z=z(\rho)$ determining a surface in
the Euclidean space with the line element
\begin{equation}                                              
  \mathrm{d}\ell^2_\mathrm{(E)} =
    \left[1+\left(\oder{z}{\rho}\right)^2 \right]\mathrm{d}\rho^2 +
    \rho^2\,\mathrm{d}\alpha^2                             \label{Euclidel}
\end{equation}
isometric to the 2-dimensional equatorial plane of the ordinary or the optical
space line element \citep{Stu-Hle:1999:ACTPS2:}
\begin{equation}                                               
  \mathrm{d}\ell^2 =
    h_{rr}\,\mathrm{d} r^2 +
    h_{\phi\phi}\,\mathrm{d}\phi^2.
\end{equation}
The azimuthal coordinates can be identified ($\alpha\equiv\phi$) which
immediately leads to $\rho^2 = h_{\phi\phi}$, and the embedding formula is
governed by the relation
\begin{equation}                                         
  \left(\oder{z}{\rho} \right)^2 =
    h_{rr} \left(\oder{r}{\rho}\right)^2 - 1.
\end{equation}
It is convenient to transfer the embedding formula into a parametric form
$z(\rho) = z(r(\rho))$ with $r$ being the parameter. Then
\begin{equation}                           
  \oder{z}{r} =
    \pm\sqrt{h_{rr} -
    \left(\oder{\rho}{r}\right)^2}.                        \label{e37}
\end{equation}
The sign in this formula is irrelevant, leading to isometric surfaces.
Because $\loder{z}{\rho} = (\loder{z}{r})(\loder{r}{\rho})$, the turning
points of the embedding diagram, giving its throats and bellies, are
determined by the condition $\mathrm{d}\rho/\mathrm{d} r = 0$.  The \ed{} can
be constructed if the reality condition $h_{rr} -
(\dif\rho/\dif r)^2 \geq 0$ holds.

\section{Examples of embedding diagrams}\label{examples}%%%%%%%%%%%%%%%%%%%

In this section we present graphic representations of embedding diagrams for
various spacetimes with their brief description. For full treatment, see the
reference provided at the beginning of each subsection.

\subsection{Schwarzschild spacetime
  with nonzero cosmological constant}\label{SdS}%//////////////////////////

The influence of cosmological constant on the structure of \SdS{} spacetime is
treated in full technical details by \citet{Stu-Hle:1999:PHYSR4:}. Putting $y
\equiv\Lambda M^2/3$, $t \to t/M$, $r \to r/M$, where $M$ is the
mass parameter and $\Lambda$ denotes the cosmological constant, the line
element in standard Schwarzschild coordinates reads
\begin{equation}
  \dif s^2
    = -\left(1 - \frac{2}{r} - yr^2\right)\dif t^2
    + \left(1 - \frac{2}{r} - yr^2\right)^{-1}\dif r^2
    + r^2 (\dif\theta^2 + \sin^2 \theta\,\dif\phi^2)\,.
\end{equation}

Embedding diagrams of the ordinary geometry are given by the formula $z=z(r)$,
which can be obtained by integrating the relation
\begin{equation}
  \frac{\dif z}{\dif r}
    = \left(\frac{2+yr^3}{r-2-yr^3}\right)^{1/2}\,.  \label{tpa010embedord}
\end{equation}
In the case of \SdS\ spacetimes, the embedding can be constructed for complete
static regions between the black-hole ($r_\mathrm{h}$) and cosmological
($r_\mathrm{c}$) horizons. Recall that the static region exists for $y\leq
1/27$ only.  In the case of \SadS\ spacetimes, the static region extends from
the black-hole horizon to infinity. However, we can see directly from
Eq.\,(\ref{tpa010embedord}) that the embedding diagrams of the ordinary space
can be constructed in a limited part of the static region, located between the
black-hole horizon $r_\mathrm{h}$ and $r_{\rm e(ord)}=(-2/y)^{1/3}$.

Embedding diagrams of optical reference geometry are given by parametric
formulas $z=z(r), \rho=\rho(r)$:
\begin{equation}                                                 
  \oder{z}{r}
    = \frac{r}{r-2-yr^3}
      \left[\frac{4r-9-yr^4}{r(r-2-yr^3)}
     \right]^{1/2}\,,\quad
  \rho = r \left(1-\frac{2}{r}-yr^2\right)^{-1/2}\,.                                 \label{tpa010embedopt}
\end{equation}
The embedding formula $z=z(\rho)$ can then be constructed by a numerical
procedure. Further, it can be shown \citep{Stu-Hle:1999:PHYSR4:} that
`turning radii' of the embedding diagrams are given by the condition
$\dif\rho/\dif r =0$. Since
\begin{equation}
  \oder{\rho}{r} = \left(1-\frac{3}{r}\right)
    \left(1-\frac{2}{r}-yr^2\right)^{-3/2},
\end{equation}
we can see that the turning radius determining a throat of the embedding
diagram of the optical geometry is located just at $r=3$, corresponding to
the radius of the photon circular orbit; it is exactly the same result as
that obtained in the pure \Sch\ case. The radius of the photon circular
orbit is important from the dynamical point of view, because the
centrifugal force related to the optical geometry reverses its sign
there~\citep{Abr-Pra:1990:MONNR:,Stu:1990:BULAI:}. Above the photon circular
orbit, the dynamics is qualitatively Newtonian with the centrifugal force
directed towards increasing $r$. However, at $r=3$, the centrifugal force
vanish, and at $r<3M$ it is directed towards decreasing $r$.  The photon
circular orbit, the throat of the embedding diagram of the optical geometry
($\dif\rho/\dif r=0$), and vanishing centrifugal force, all appear at the
radius $r=3$.

\begin{figure}[b]
\begin{minipage}{.29\linewidth}
\centering
\includegraphics[width=\linewidth]{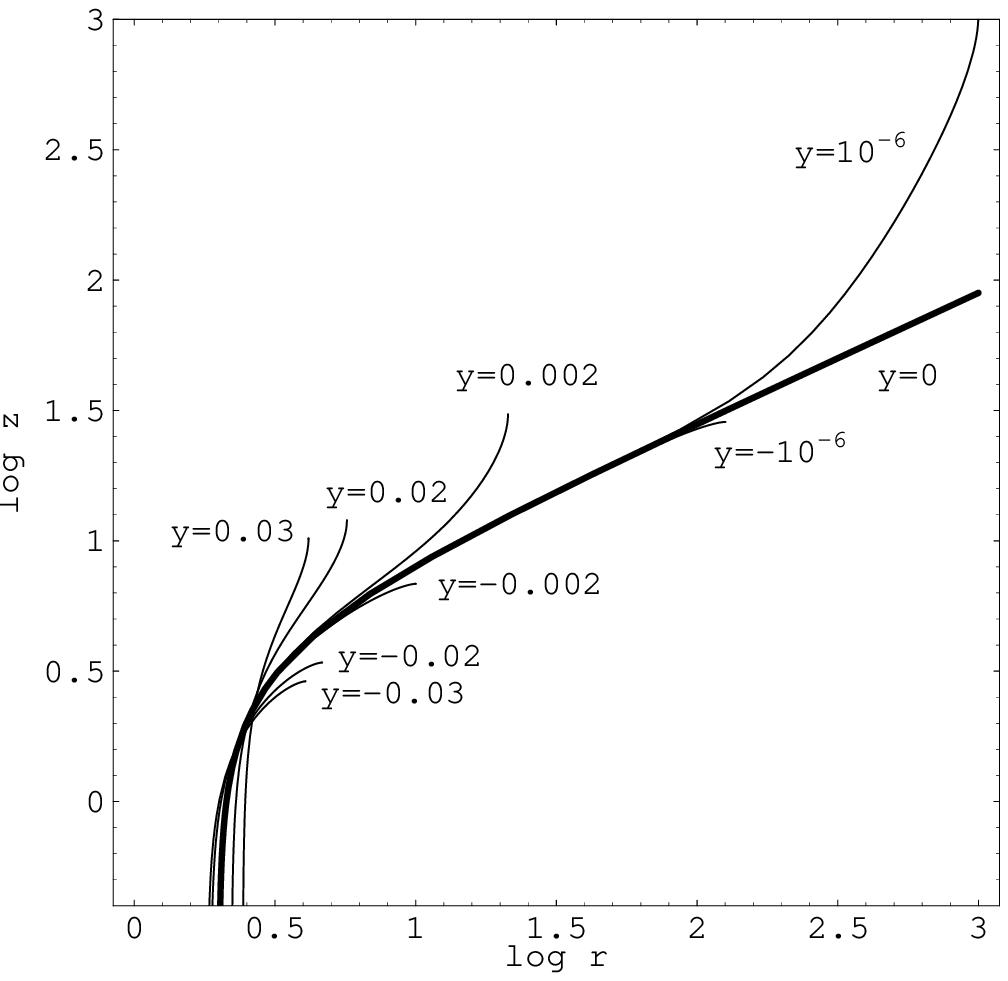}
\end{minipage}\enspace%
\begin{minipage}{.29\linewidth}
\centering
\includegraphics[width=\linewidth]{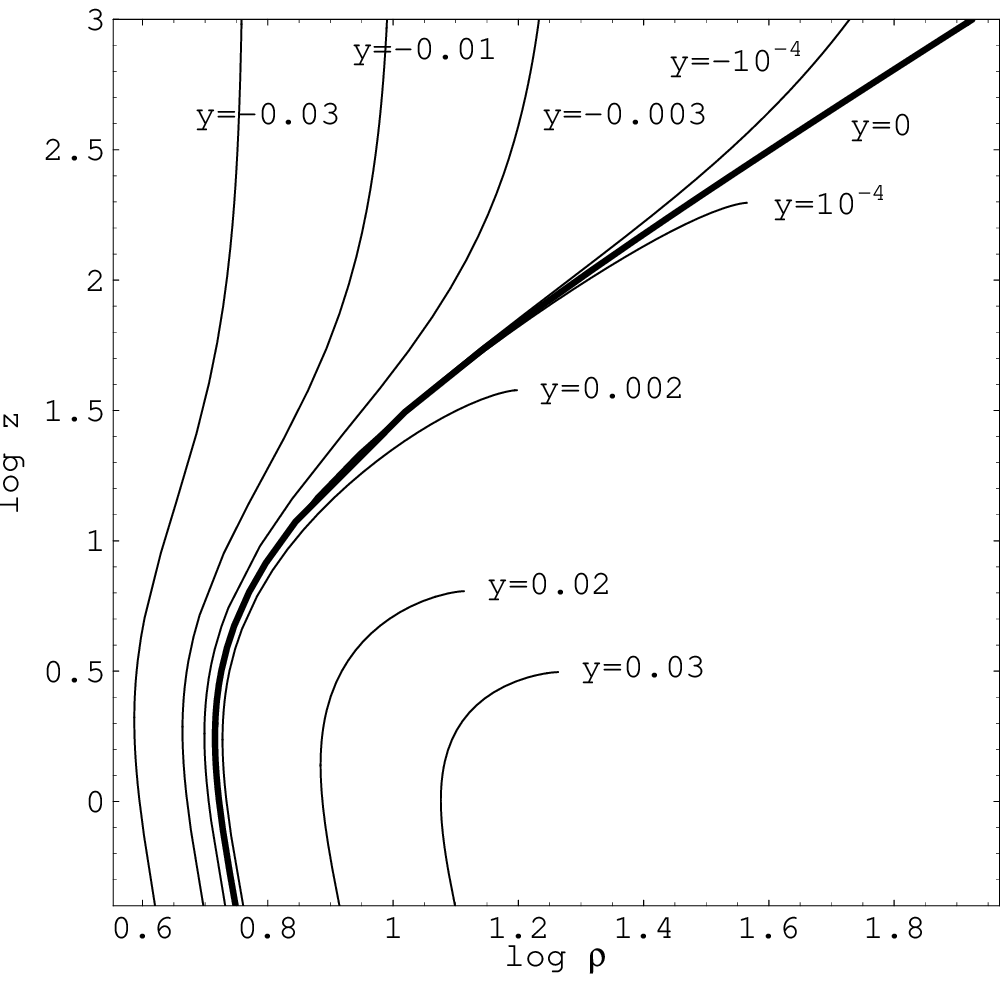}
\end{minipage}\enspace\enspace\enspace%
\begin{minipage}{.37\linewidth}
\centering
\includegraphics[width=\linewidth]{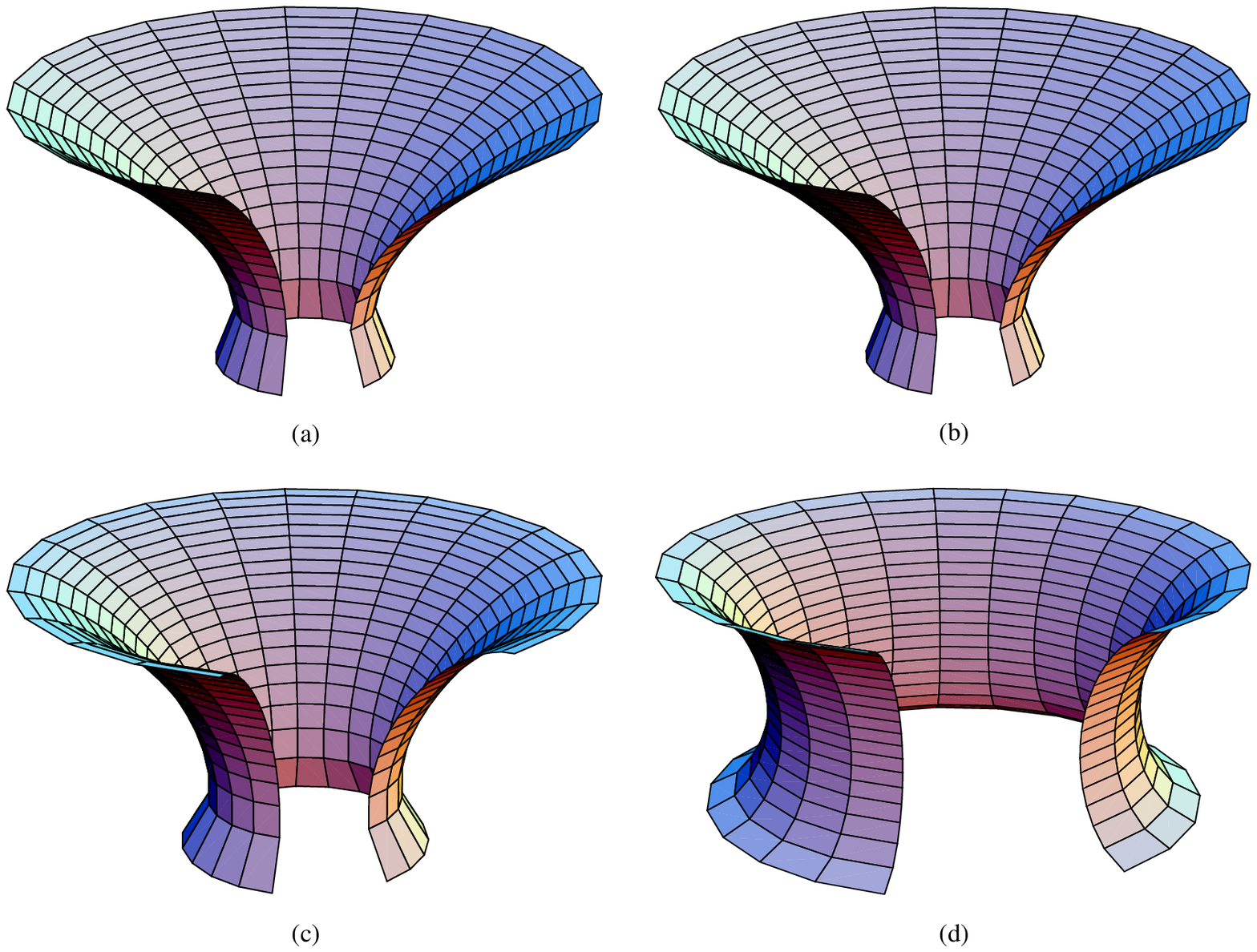}
\end{minipage}
\caption{\label{tpa010f10}\textit{Left panel:} Qualitative features of the
  embedding diagrams of the ordinary geometry of the \SdS\ and \SadS\ 
  spacetimes in a log-log diagram. One can immediately see how the diagrams
  with $y\neq 0$ `peel off' the pure \Sch\ diagram ($y=0$, bold curve). All
  sections with $y\neq 0$ are complete (i.e., the maximum that can be embedded
  into Euclidean space is shown), except uninteresting lower parts of the
  throats. The diagrams clearly indicate modifications of the spacetime
  structure caused by the presence of a cosmological constant.  \textit{Middle
    panel:} Same as left panel but for optical geometry. \textit{Right panel:}
  Example of embedding diagrams of the optical reference geometry of the \SdS\ 
  spacetimes. The pure \Sch\ case is taken for comparison in (a).  The
  diagrams are given for (b)~$y=10^{-6}$, (c)~$y=0.002$, and (d)~$y=0.03$.
  They are similar to the pure \Sch\ case, because the region near the
  cosmological horizon being of highly different character is `cut off' by the
  limit of embeddability. \takenfrom{Stu-Hle:1999:PHYSR4:}}
\end{figure}

In the case of \SdS\ spacetimes containing the static region, the
embeddability of the optical geometry is restricted both from below, and from
above. Using a numerical procedure, the embedding diagrams are constructed for
the same values of $y$ as in the case of the ordinary geometry.

In the case of \SadS\ spacetimes, the embeddability of the optical geometry is
restricted from below again; with $y \to -\infty$ the limit shifts to $r \to
0$, along with the radius of the black-hole horizon. The embedding diagrams
are constructed by the numerical procedure for the same values of $y$ as for
the ordinary space.  These diagrams
have a special property, not present for the embedding diagrams in the other
cases.  Namely, they cover whole the asymptotic part of the \SadS\ spacetime,
but in a restricted part of the Euclidean space. This is clear from the
asymptotic behavior of $\rho(r)$.  For $r \to + \infty$, there is $\rho \sim
(-y)^{-1/2}$. Clearly, with decreasing attractive cosmological constant the
embedding diagram is deformed with increasing intensity. The circles of $r =
\mathrm{const}$ are concentrated with an increasing density around $\rho =
(-y)^{-1/2}$ as $r \to \infty$.

Basic features of the embedding diagrams of the \SdS{} spacetimes are
illustrated in Figure~\ref{tpa010f10}.

\subsection{Interior uniform-density Schwarzschild spacetime with
  nonzero cosmological constant}\label{iscc}%////////////////////////////
  
The influence of the cosmological constant on the structure of interior
uniform-density Schwarzschild spacetime is treated in full technical details
by \citet{Stu-etal:2001:PHYSR4:}.

\begin{figure}[h]
\begin{minipage}{.46\linewidth}
\centering
\includegraphics[width=\linewidth]{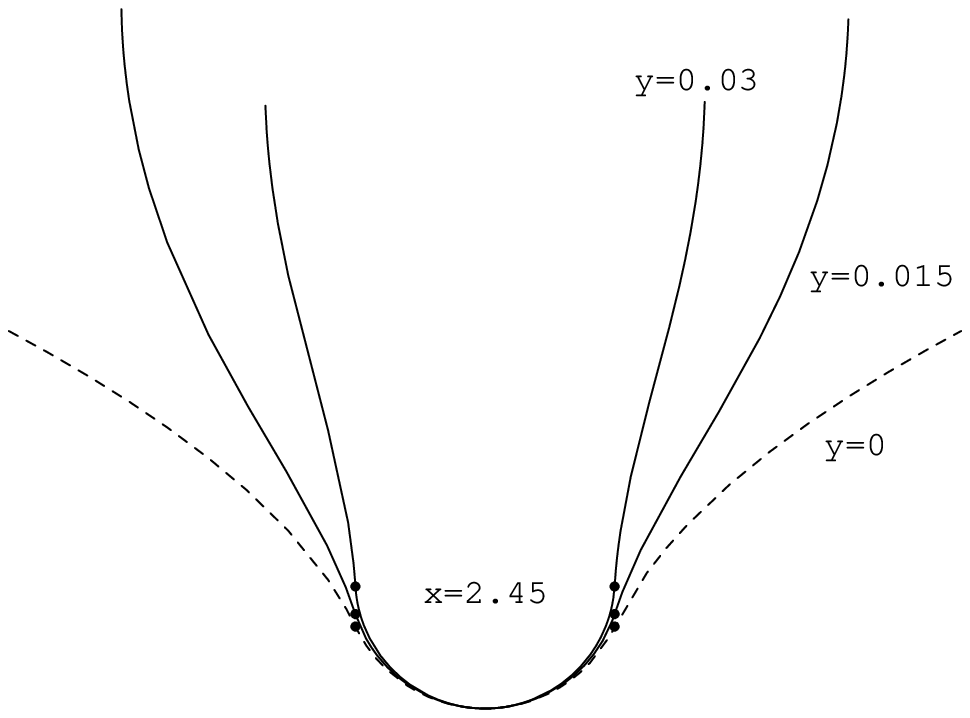}
\par
Ordinary geometry
\end{minipage}\hspace{6em}%
\begin{minipage}{.28\linewidth}
\centering
\includegraphics[width=\linewidth]{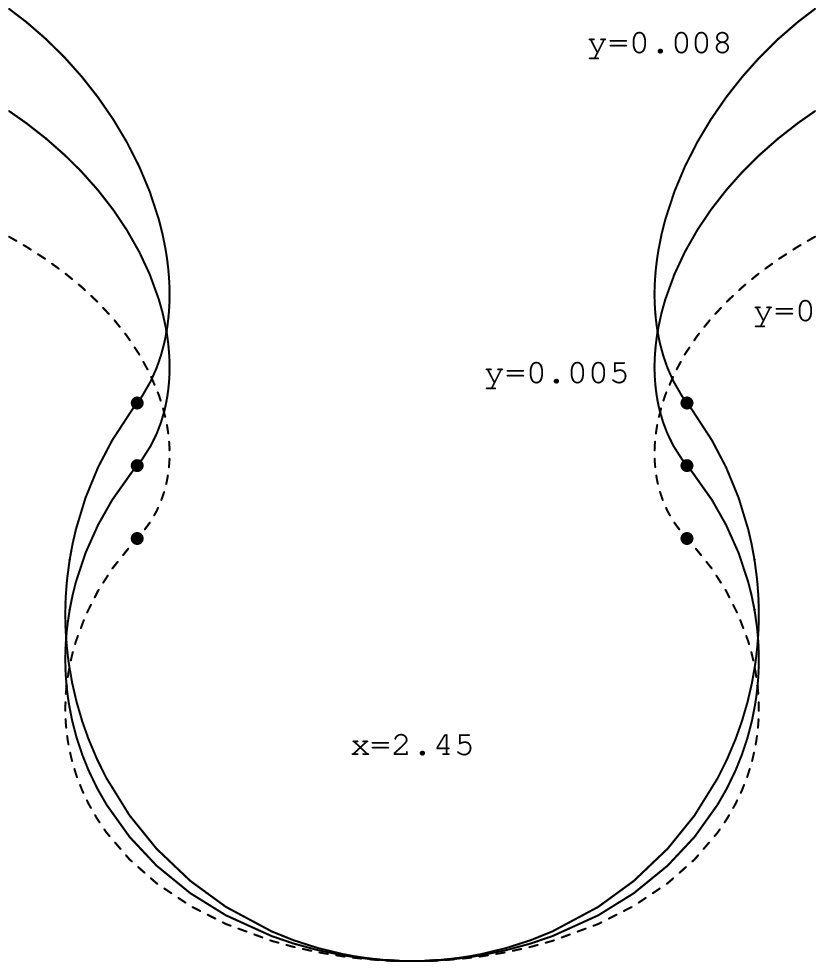}
\par
Optical geometry
\end{minipage}
\caption{\label{isccr}Axial sections of the Schwarzschild--de Sitter
  ($\Lambda>0$) case. The dashed curve represents pure \Sch\ spacetime, the
  dots mark the position of the edge of the internal spacetime (i.e.,
  $r=R=xM$). The embedding diagrams of the optical reference geometry show the
  existence of internal circular null geodesics which play a key role for the
  possibility of neutrino trapping.}
\end{figure}

The line element in standard Schwarzschild coordinates reads
\begin{equation}
  \dif s^2 =
    - \eto{2\Phi(r)}\,\dif t^2
    + \eto{2 \Psi(r)}\,\dif r^2
    + r^2 (\dif\theta^2
    + \sin^2\theta\,\dif\varphi^2)\,,
\end{equation}
where
\begin{equation}
  \eto{\Psi(r)} = \left(1-\frac{r^2}{a^2}\right)^{-1/2}\,,\;
  \eto{\Phi(r)} = A\eto{-\Psi(R)} - B\eto{-\Psi(r)}\,,\;
  \frac{1}{a^2} = \frac{2M}{R^3} + \frac{\Lambda}{3}\,,\;
  A = \frac{9M}{6M + \Lambda R^3}\,,\;
  B = \frac{3M - \Lambda R^3}{6M + \Lambda R^3}\,.
\end{equation}
Here $M$ denotes total mass, $\rho$ (uniform) mass density and $R$ external
radius of the configuration; $\Lambda$ is the cosmological constant. Examples
of embedding diagrams (for repulsive cosmological constant) are shown in
Figure~\ref{isccr}.\\[2ex]
\null

\subsection{\RN{} spacetime with nonzero cosmological constant}\label{rncc}

This section is based on \citep{Stu-Hle:2002:ACTPS2:}, where full-detail
treatment can be found.  Putting $y \equiv \tfrac{1}{3} \Lambda M^2$, $e
\equiv Q/M$, $t \to t/M$, $r \to r/M$, where $M$ is the mass parameter, $Q$ is
the electric charge and $\Lambda$ is the cosmological constant, the line
element in standard Schwarzschild coordinates reads
\begin{equation}
  \dif s^2 =
    -\left(
      1- \frac{2}{r}+ \frac{e^2}{r^2} - y{r^2}
     \right)\dif t^2
    + \left(
      1- \frac{2}{r}+ \frac{e^2}{r^2} - y{r^2}
     \right)^{-1} \dif r^2
    +r^2\left(\dif\theta^2+\sin^2\theta\,\dif\phi^2\right)\,.
\end{equation}

\begin{figure}[t]
\newlength{\adhocdim}
\setlength{\adhocdim}{5.12in}
\null
\begin{picture}(0,105)(65,-85)
\put(40,-50){\includegraphics[width=.4\adhocdim]{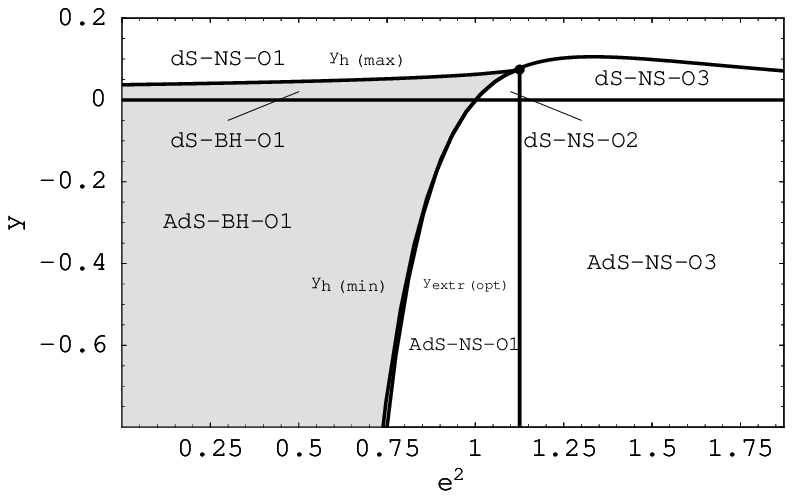}}

\put(0,-30){\includegraphics[width=.3\adhocdim]{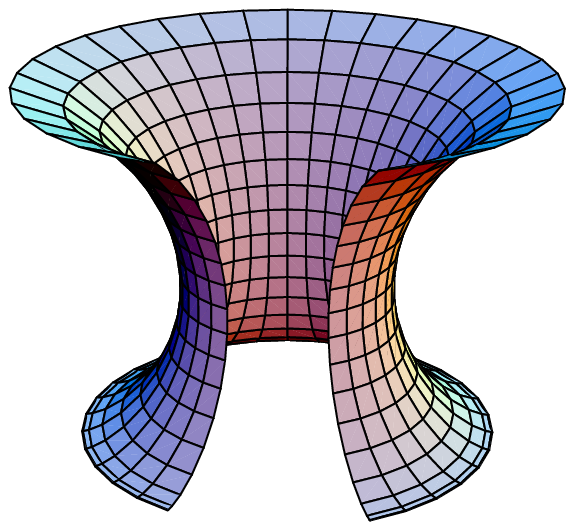}}
\put(40,-8){\vector(2,-1){30}}

\put(0,-80){\includegraphics[width=.3\adhocdim]{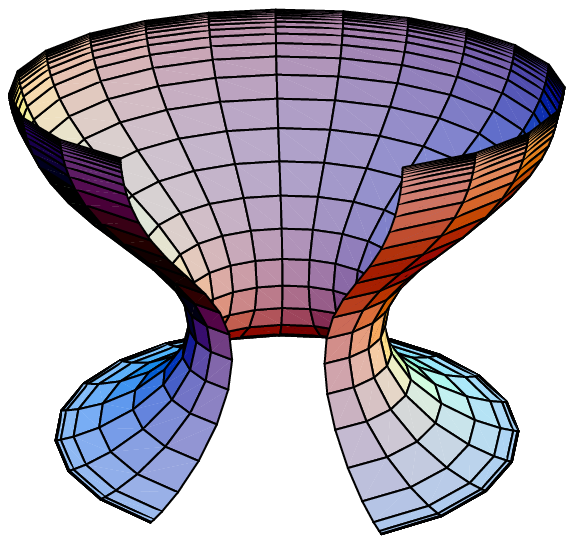}}
\put(40,-40){\vector(2,1){30}}

\put(90,-30){\includegraphics[width=.3\adhocdim]{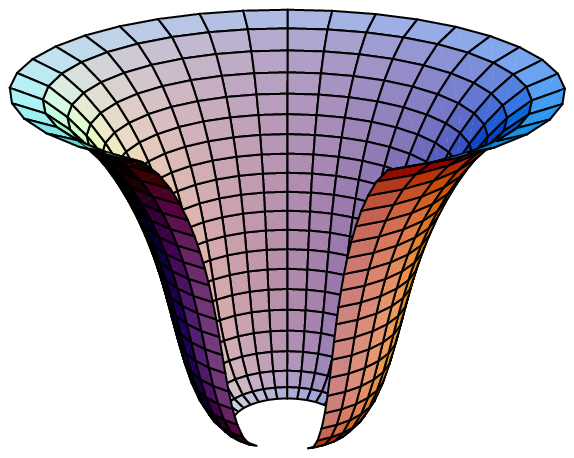}}
\put(89,-9){\vector(-1,-1){14}}

\put(90,-80){\includegraphics[width=.3\adhocdim]{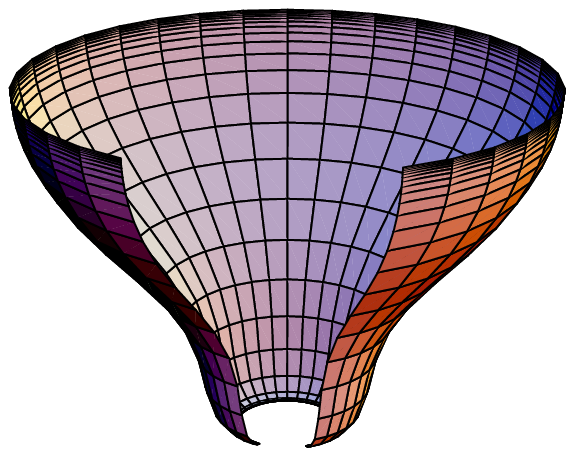}}
\put(97,-47){\vector(-1,1){22}}

\put(48,-16){\includegraphics[width=.3\adhocdim]{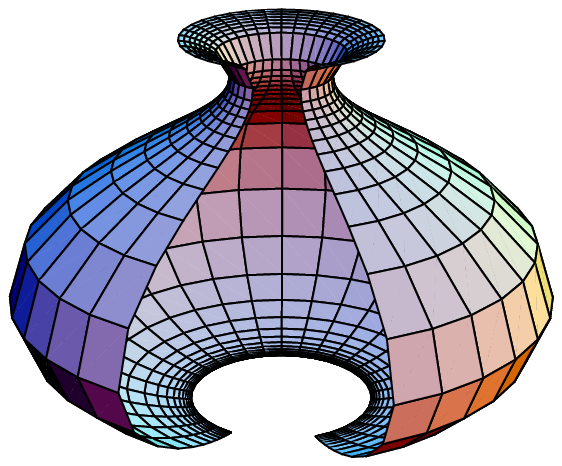}}
\put(73,-17){\vector(0,-1){6}}

\put(48,-86){\includegraphics[width=.28\adhocdim]{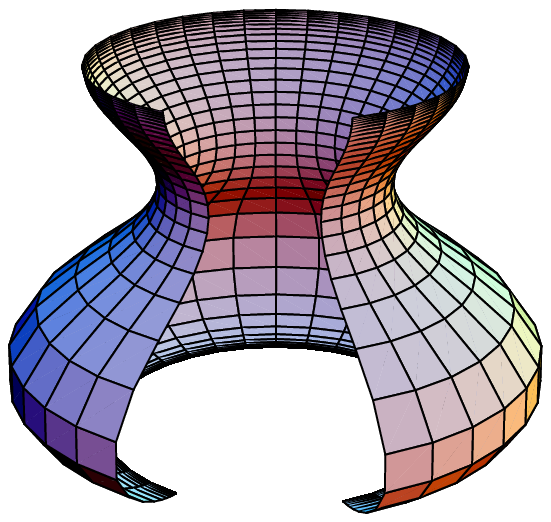}}
\put(71,-51){\vector(0,1){26}}

\put(100,-42){not embeddable}
\put(98,-39.5){\vector(-1,2){10}}
\end{picture}
\caption{\label{rnccfig}An example classification of \RN{} spacetimes
  according to properties of embedding diagrams of optical reference geometry.
  \takenfrom{Stu-Hle:2002:ACTPS2:}}
\end{figure}

An example of classification based on properties of embedding diagrams of
optical reference geometry is in Figure~\ref{rnccfig}.

\subsection{Ernst spacetime}\label{Ernst}%/////////////////////////////////

The static Ernst spacetime~\citep{Ern:1976:JMATP4:,Stu-Hle:1999:CLAQG:} is
the only exact solution of Einstein's equations known to represent the
spacetime of a spherically symmetric massive body or black hole of mass $M$
immersed in an otherwise homogeneous magnetic field. If the magnetic field
disappears, the geometry simplifies to the \Sch{} geometry.  Therefore,
sometimes the Ernst spacetime is called magnetized \Sch{} spacetime.
Usually it is believed that for extended structures like galaxies both the
effects of general relativity and the role of a magnetic field can be
ignored.  However, in the case of active galactic nuclei with a huge
central black hole and an important magnetic field, the Ernst spacetime can
represent some relevant properties of the galactic structures. Therefore,
it could even be astrophysically important to discuss and illustrate basic
properties of the Ernst spacetime.

The Ernst spacetime has the important property that it is not
asymptotically flat. Far from the black hole, the spacetime is closely
related to Melvin's magnetic universe, representing a cylindrically
symmetric spacetime filled with an uniform magnetic field only. The Ernst
spacetime is axially symmetric, and its structure corresponds to the
structure of the \Sch{} spacetime only along its axis of symmetry.  Off the
axis, the differences are very spectacular, and we shall demonstrate them
for the equatorial plane, which is the symmetry plane of the spacetime.

The line element of the Ernst spacetime reads
\begin{equation}
\dif s^2 = \Lambda^2 \left[-\left(1-\frac{2M}{r}\right)\dif t^2 +
    \left(1-\frac{2M}{r}\right)^{-1}\dif r^2 + r^2\,\dif\theta^2 \right]
    + \frac{r^2 \sin^2 \theta}{\Lambda^2}\,\dif\phi^2\,,
\end{equation}
where $M\equiv M_{\mathrm{cgs}}G/c^2$ is the mass, $B\equiv
B_{\mathrm{cgs}}G^{1/2}/c^2$ is the strength of the magnetic field,
$\Lambda\equiv 1 + B^2 r^2 \sin^2 \theta$.

\begin{figure}[t]
\begin{minipage}{.30\linewidth}
\centering
\includegraphics[width=.9\linewidth]{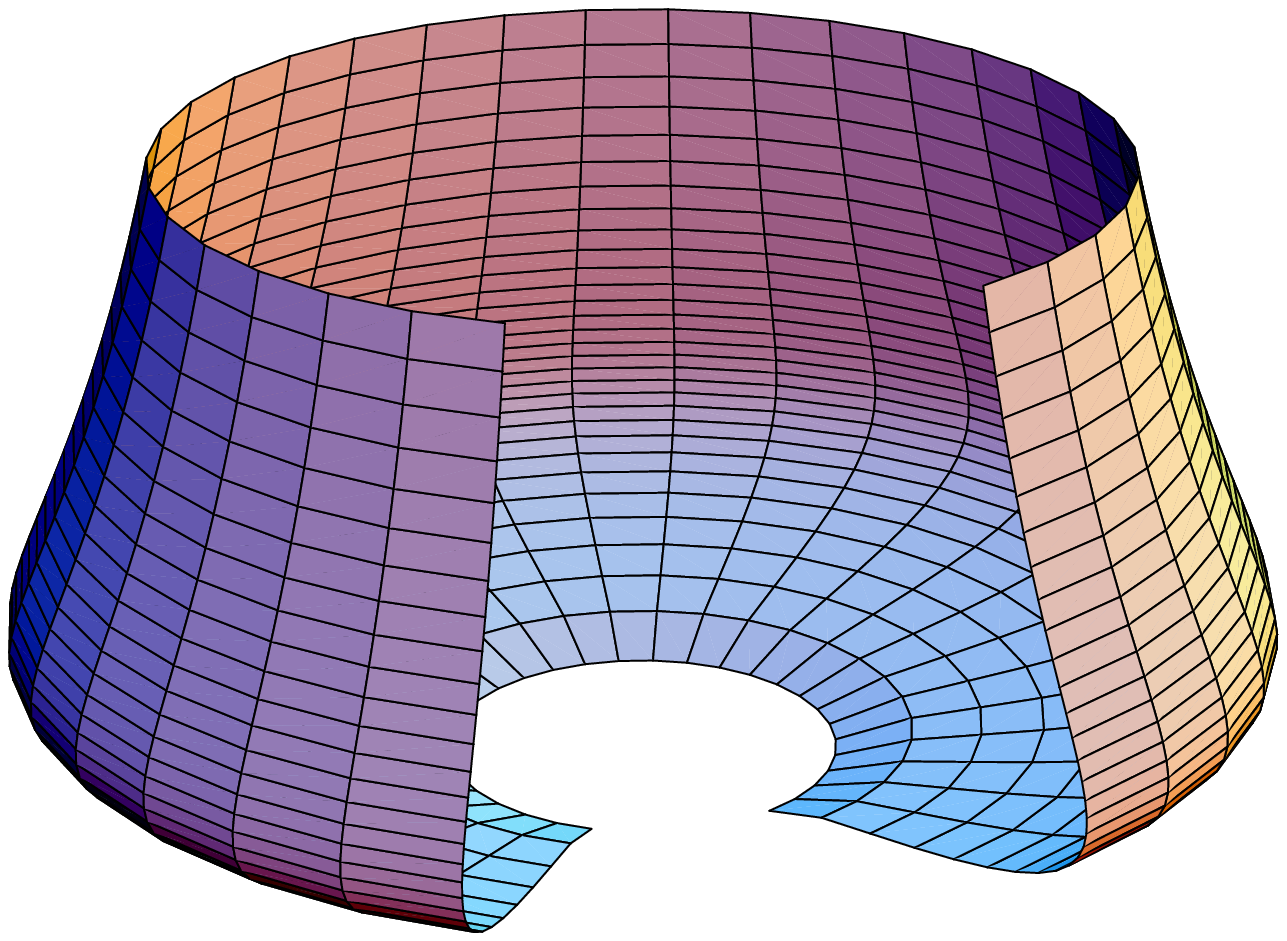}
\end{minipage}\hfill%
\begin{minipage}{.22\linewidth}
\centering
$B=0.08<B_{\mathrm{c}}$
\end{minipage}
\begin{minipage}{.30\linewidth}
\centering
\includegraphics[width=.98\linewidth]{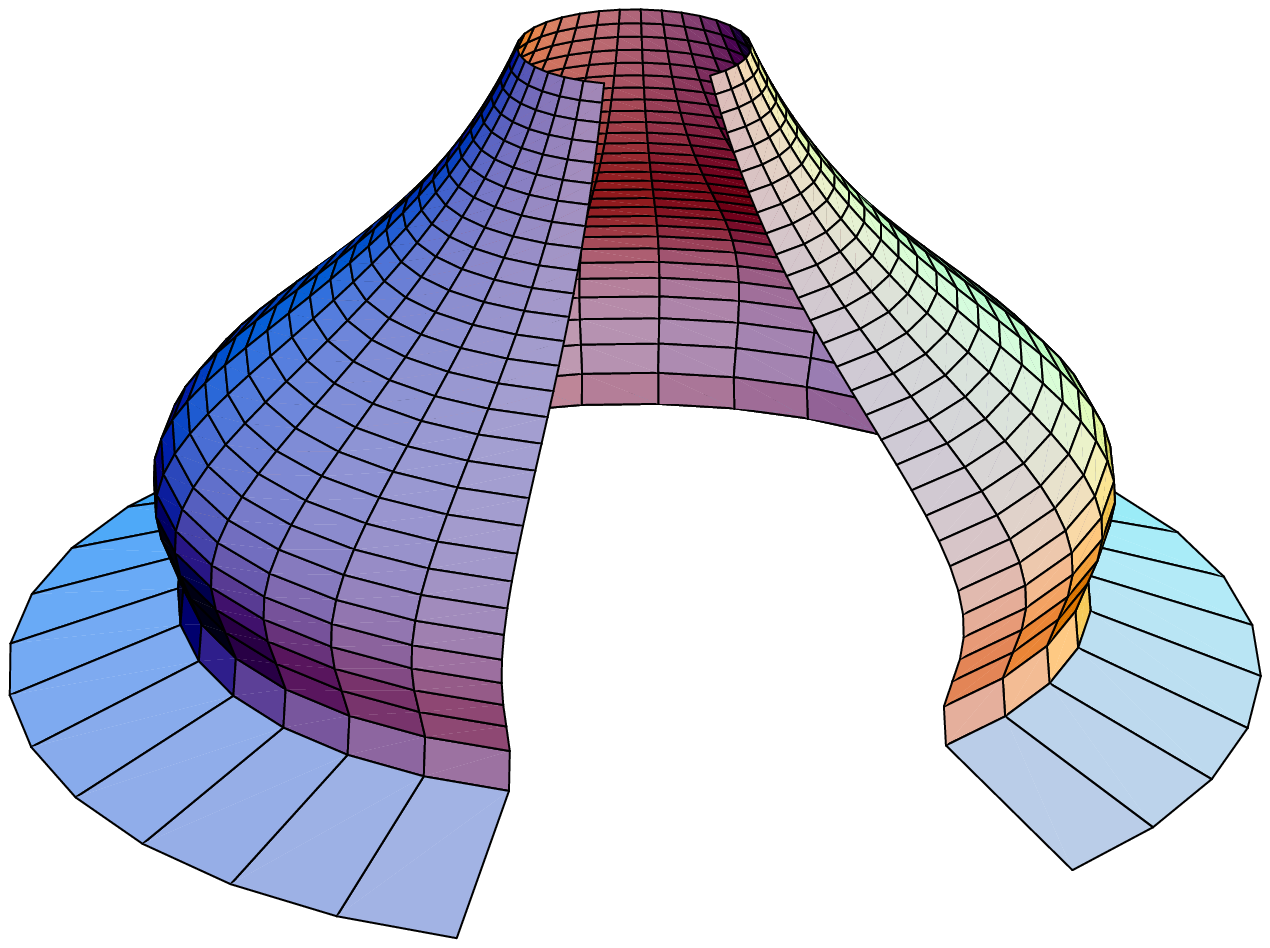}
\end{minipage}
\caption{\label{ernstfig}\textit{Left column:} ordinary geometry,
  \textit{right column:} optical geometry of Ernst spacetime. It can be proved
  \citep{Stu-Hle:1999:CLAQG:} that a critical magnetic field
  $B_{\mathrm{c}}\approx 0.0947$ exists. For $B>B_{\mathrm{c}}$, neither
  throats nor bellies and no circular photon orbits exist.  For
  $B<B_{\mathrm{c}}$, the throat and the belly develop, corresponding to the
  inner unstable and outer stable photon circular orbit.}
\end{figure}

The dimensionless product $BM\ll1$ in astrophysically realistic situations.
Really, $BM \sim 2\times 10^{-53} B_{\mathrm{cgs}}M_{\mathrm{cgs}}$.  For
example, in case of a black hole with $M = 10^9 M_{\odot}$ (typical for AGN),
the value $BM \sim 1$ corresponds to $B_{\mathrm{cgs}}\sim 10^{11}$\,Gauss,
which is unrealistic.  Some illustrative embedding diagrams are collected in
Figure~\ref{ernstfig}.

\begin{figure}[t]
\begin{minipage}{.30\linewidth}
\centering
\includegraphics[width=.9\linewidth]{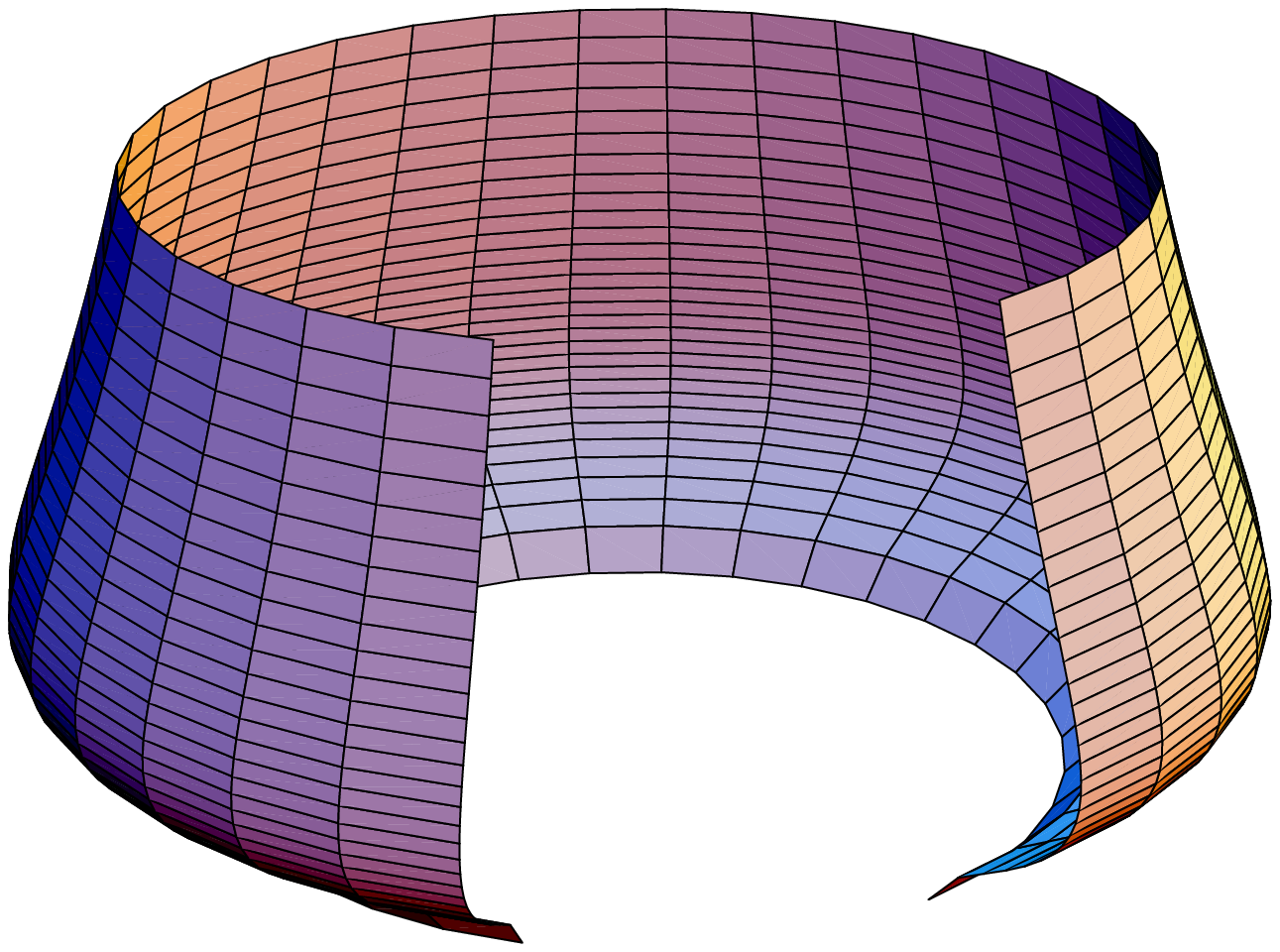}
\end{minipage}\hfill%
\begin{minipage}{.20\linewidth}
\centering
$B=0.2>B_{\mathrm{c}}$
\end{minipage}\hfill%
\begin{minipage}{.30\linewidth}
\centering
\includegraphics[width=.9\linewidth]{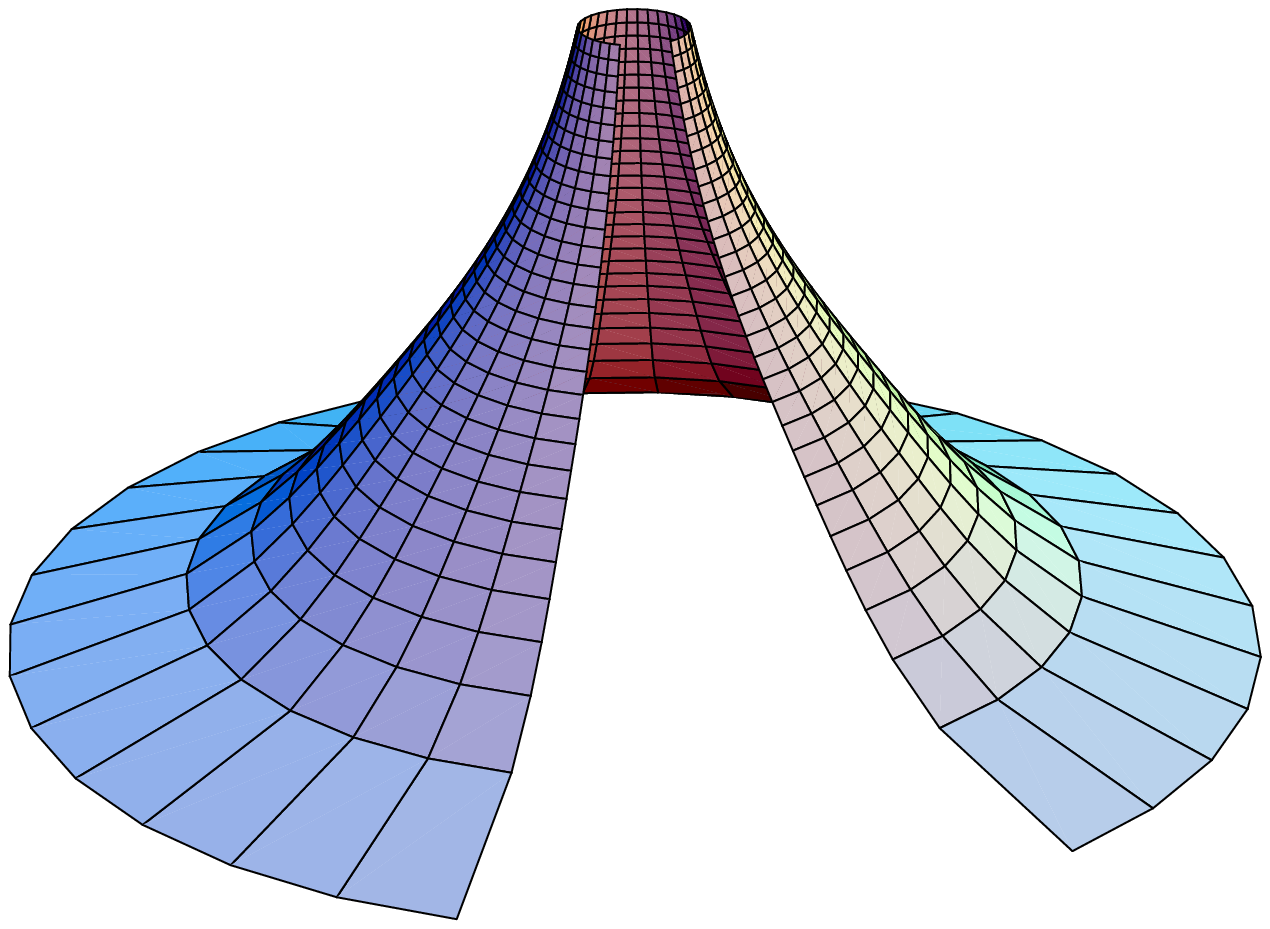}
\end{minipage}
\end{figure}

\subsection{\KN{} spacetimes}\label{KNed}%/////////////////////////////////

We focus on the results concerning the optical reference geometry
\citep{Stu-Hle-Jur:2000:CLAQG:}.  It can be shown
\citep{Stu-Hle-Jur:2000:CLAQG:} that the turning points of the embedding
diagrams are really located at the radii where the centrifugal force vanishes
and changes sign. Thus, we can conclude that it is exactly this property of
optical geometry embeddings for the vacuum spherically symmetric spacetimes
(see
\citep{Abr-Car-Las:1988:GENRG2:,Kri-Son-Abr:1998:GENRG2:,Stu-Hle:1999:PHYSR4:})
survives in the \KN{} spacetimes.  However, photon circular orbits are
displaced from the radii corresponding to the turning points of the embedding
diagrams.

\begin{figure}[b]
\begin{minipage}[b]{.19\linewidth}
\centering
\includegraphics[width=.95\linewidth]{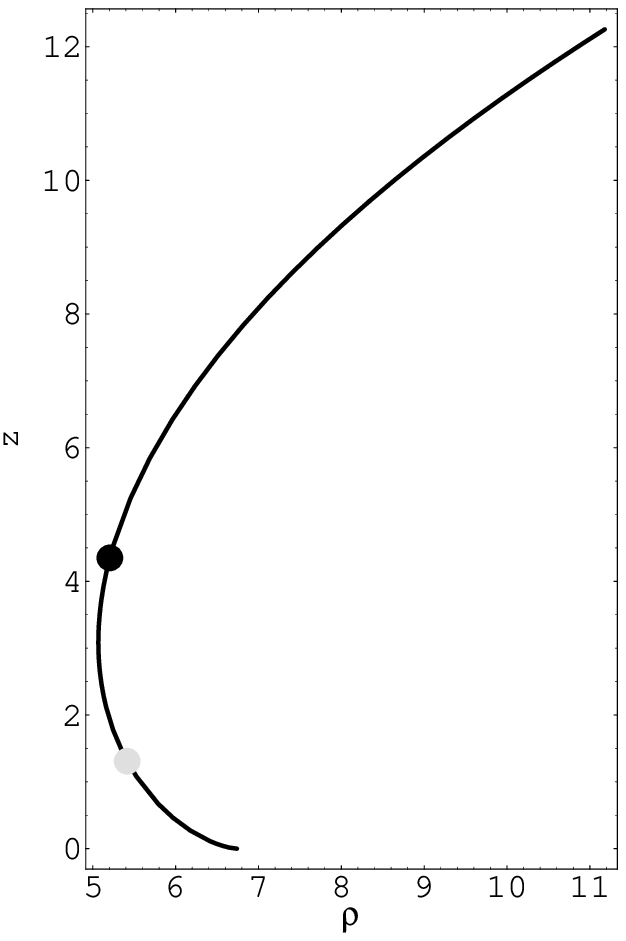}
\par {\small (a)}
\end{minipage}\hfill%
\begin{minipage}[b]{.32\linewidth}
\centering
\includegraphics[width=.9\linewidth]{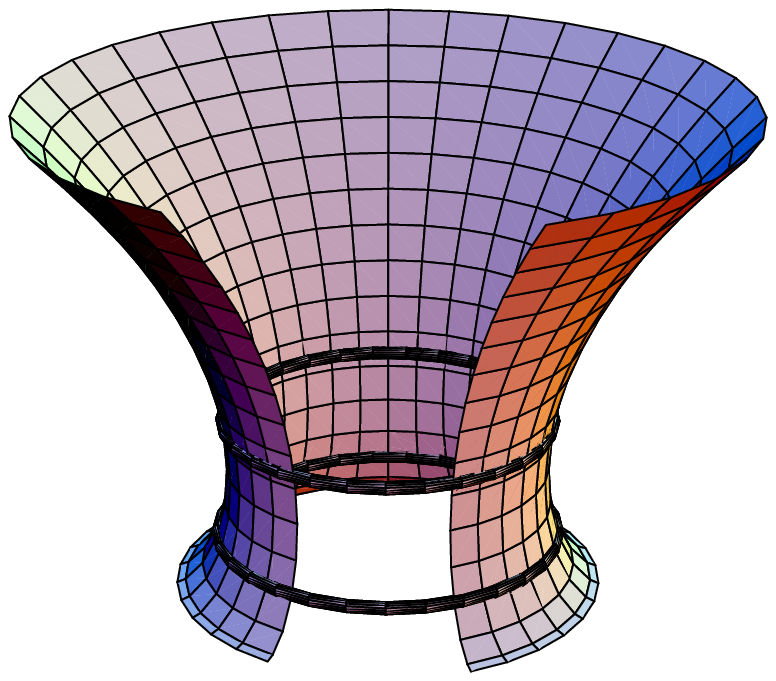}
\par {\small (b)}
\end{minipage}\hfill%
\begin{minipage}[b]{.15\linewidth}
\centering
\includegraphics[width=.9\linewidth]{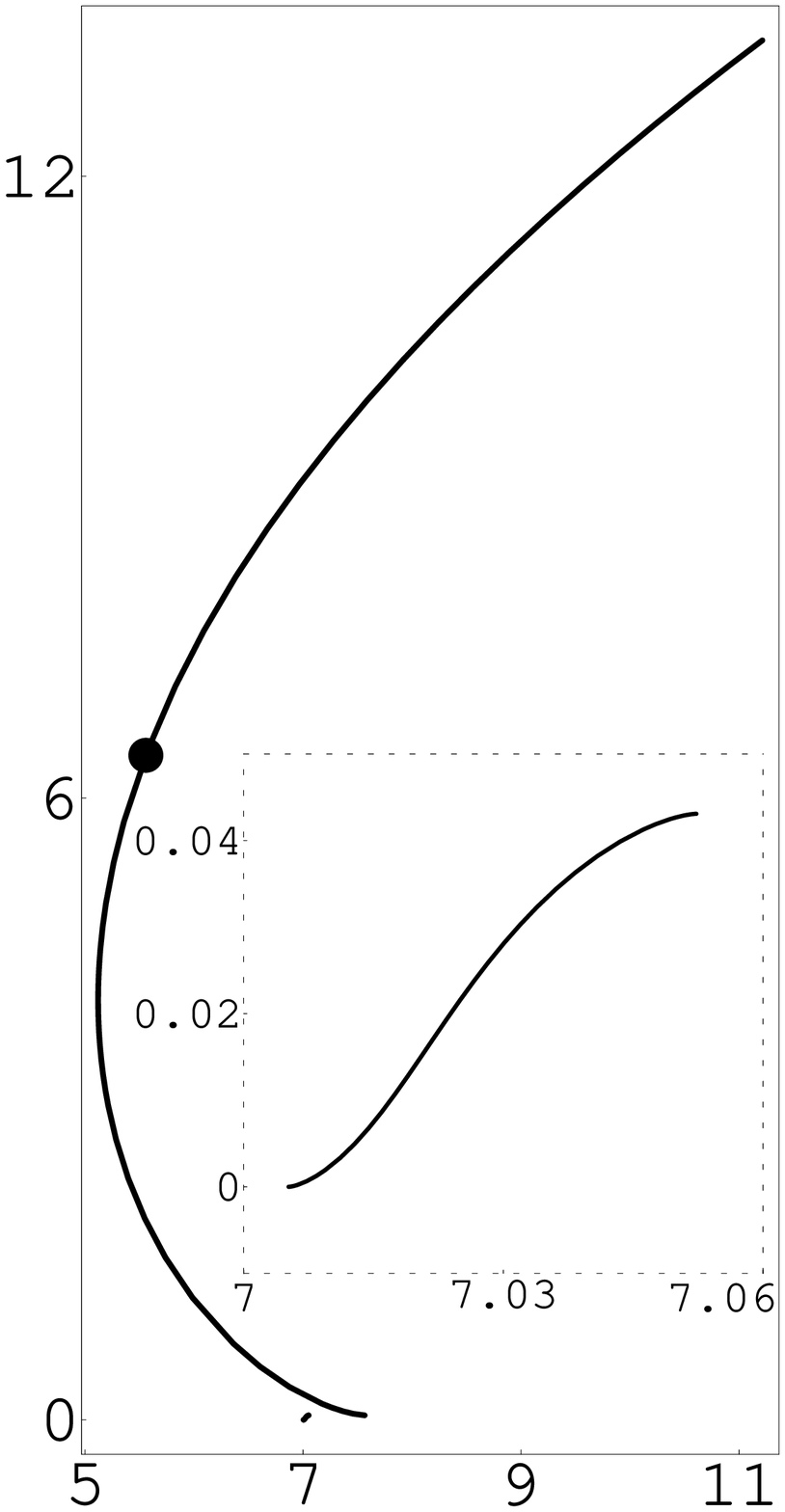}
\par {\small (c)}
\end{minipage}\hfill%
\begin{minipage}[b]{.14\linewidth}
\centering
\includegraphics[width=.962\linewidth]{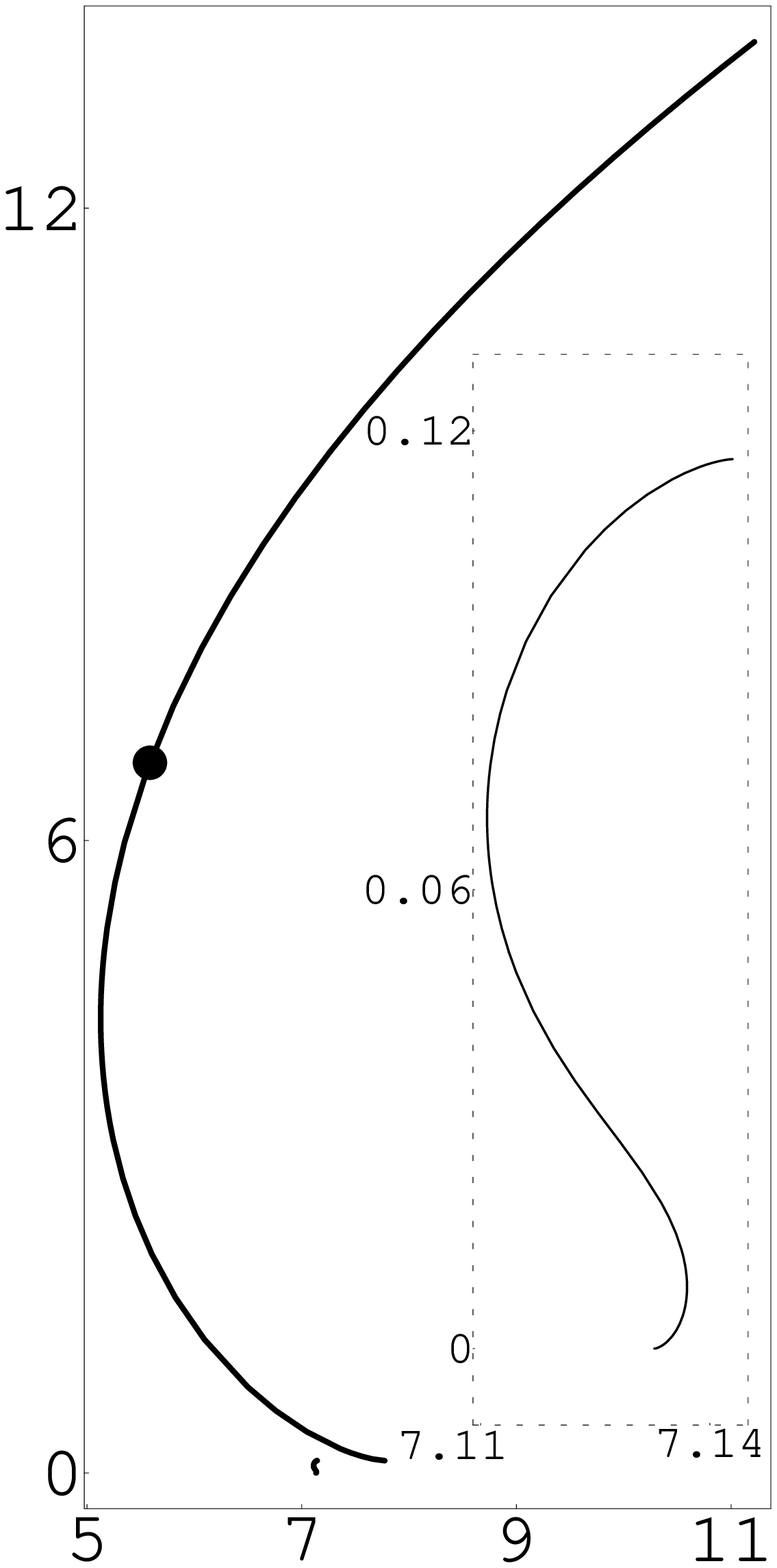}
\par {\small (d)}
\end{minipage}\hfill%
\begin{minipage}[b]{.14\linewidth}
\centering
\includegraphics[width=.932\linewidth]{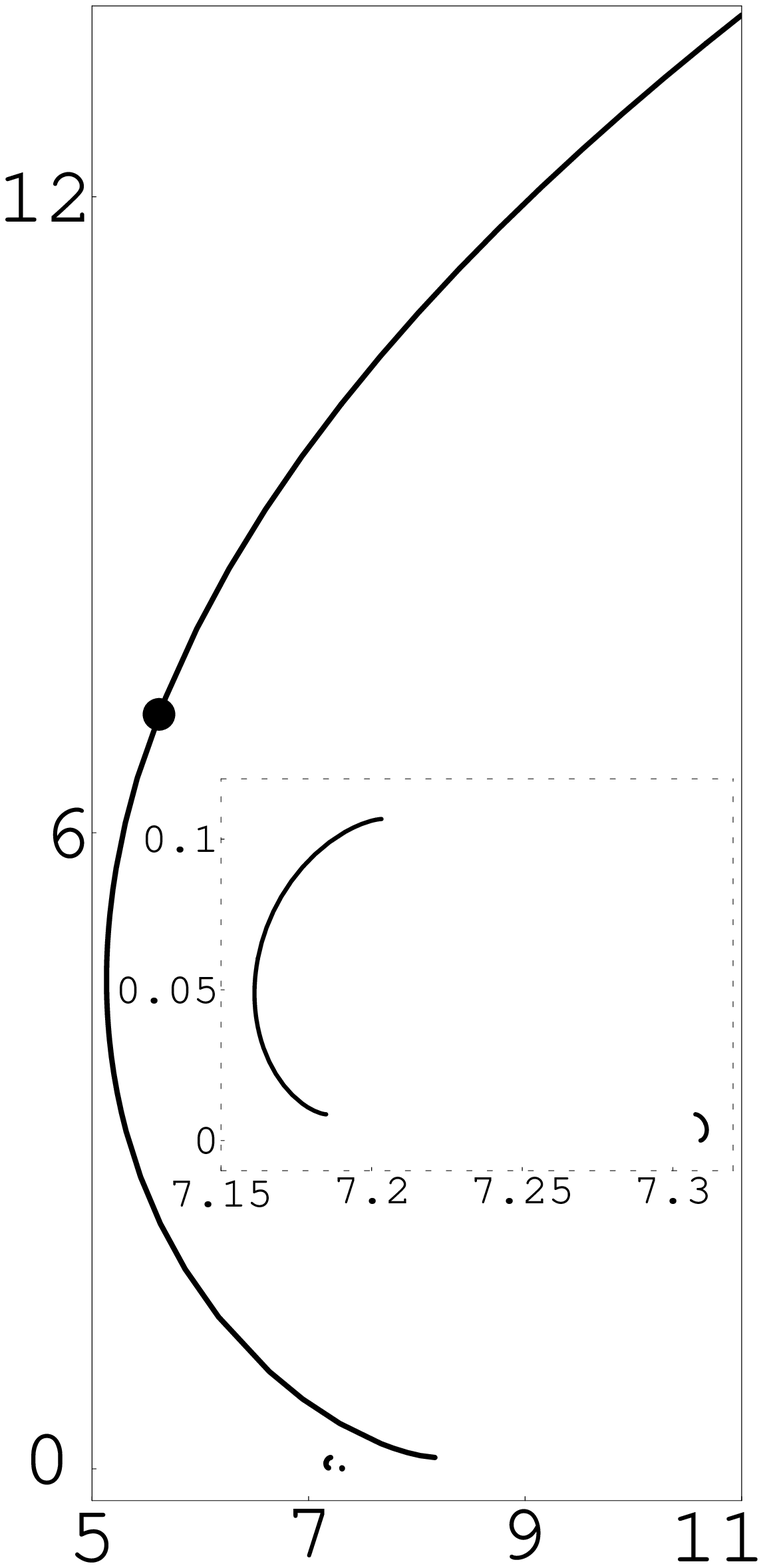}
\par {\small (e)}
\end{minipage}
\caption{\label{KNed-bh1}(a), (b): Embedding diagram of the
  \KN{} black holes of the type BH$_1$, constructed for $a^2=0.16$,
  $e^2=0.16$. The rings in the 3D diagram represent photon circular orbits.
  Both corotating (gray spot in 2D diagram) and counterrotating (black spot in
  2D diagram) orbits are displaced from the throat of the diagram where the
  centrifugal force vanishes. This is a general property of the rotating
  backgrounds. (c)--(e): Embedding diagrams of the
  black holes classes BH$_2$\nd BH$_4$.  (c)~Class BH$_2$ ($a^2=0.7$,
  $e^2=0.16$); (d)~class BH$_3$ ($a^2=0.75$, $e^2=0.16$); (e)~class BH$_4$
  ($a^2=0.81$, $e^2=0.16$).  \takenfrom{Stu-Hle-Jur:2000:CLAQG:}}
\end{figure}

After a straightforward, but rather tedious, discussion, the full
classification can be made according to the properties of the embedding
diagrams (see \citep{Stu-Hle-Jur:2000:CLAQG:}).

Examples of typical black-hole \eds{} of classes labeled in
\citet{Stu-Hle-Jur:2000:CLAQG:} as BH$_1$\nd BH$_4$ are presented in
Figure~\ref{KNed-bh1}.

\section{Concluding remarks}\label{conclusion}%%%%%%%%%%%%%%%%%%%%%%%%%%%%%

Embedding diagrams of the optical geometry give an important tool of
visualization and clarification of the dynamical behavior of test particles
moving along equatorial circular orbits: we imagine that the motion is
constrained to the surface $z(\rho)$ (see \citep{Kri-Son-Abr:1998:GENRG2:}). The shape of the surface $z(\rho)$
is directly related to the centrifugal acceleration. Within the upward
sloping areas of the embedding diagram, the centrifugal acceleration points
towards increasing values of $r$ and the dynamics of test particles has an
essentially Newtonian character. However, within the downward sloping areas
of the embedding diagrams, the centrifugal acceleration has a radically
non-Newtonian character as it points towards decreasing values of $r$.
Such a kind of behavior appears where the diagrams have a throat or a
belly. At the turning points of the diagram, the centrifugal acceleration
vanishes and changes its sign.
 
%% The black-hole backgrounds have unified character above the event horizon
%% --~there exists a throat of the embedding diagram indicating a change of sign
%% of the centrifugal force nearby the event horizon. On the other hand, the
%% naked-singularity spacetimes give a wide variety of behavior of the embeddings
%% and centrifugal forces, ranging from the simple `Newtonian' backgrounds with
%% no change of sign of centrifugal force, to the very complicated backgrounds
%% where the sign is changed four times in the \KN{} spacetimes.

\begin{theacknowledgments}
  The present work was supported by the grant MSM~4781305903.
\end{theacknowledgments}

\bibliographystyle{aipproc}
\bibliography{\jobname}

\begin{thebibliography}{26}
\expandafter\ifx\csname natexlab\endcsname\relax\def\natexlab#1{#1}\fi
\providecommand{\enquote}[1]{``#1''}
\expandafter\ifx\csname url\endcsname\relax
  \def\url#1{\texttt{#1}}\fi
\expandafter\ifx\csname urlprefix\endcsname\relax\def\urlprefix{URL }\fi
\providecommand{\eprint}[2][]{\url{#2}}

\bibitem[Stuchl{\'{\i}}k and
  Hled{\'{\i}}k(1999{\natexlab{a}})]{Stu-Hle:1999:PHYSR4:}
Z.~Stuchl{\'{\i}}k, and S.~Hled{\'{\i}}k, \emph{Phys. Rev. D} \textbf{60},
  044006 (15~pages) (1999{\natexlab{a}}).

\bibitem[Stuchl{\'{\i}}k et~al.(2001)]{Stu-etal:2001:PHYSR4:}
Z.~Stuchl{\'{\i}}k, S.~Hled{\'{\i}}k, J.~{\v{S}}olt\'es, and E.~{\O}stgaard,
  \emph{Phys. Rev. D} \textbf{64}, 044004 (17~pages) (2001).

\bibitem[Kristiansson et~al.(1998)]{Kri-Son-Abr:1998:GENRG2:}
S.~Kristiansson, S.~Sonego, and M.~A. Abramowicz, \emph{Gen. Relativity
  Gravitation} \textbf{30}, 275--288 (1998).

\bibitem[Stuchl{\'{\i}}k and
  Hled{\'{\i}}k(2002{\natexlab{a}})]{Stu-Hle:2001:FPFSU:PropRNdS}
Z.~Stuchl{\'{\i}}k, and S.~Hled{\'{\i}}k, {Properties of the
  Reissner--Nordstr{\"o}m spacetimes with a nonzero cosmological constant}
  (2002{\natexlab{a}}), unpublished, preprint TPA 003/Vol.~2, 2001.

\bibitem[Stuchl{\'{\i}}k and
  Hled{\'{\i}}k(1999{\natexlab{b}})]{Stu-Hle:1999:CLAQG:}
Z.~Stuchl{\'{\i}}k, and S.~Hled{\'{\i}}k, \emph{Classical Quantum Gravity}
  \textbf{16}, 1377--1387 (1999{\natexlab{b}}).

\bibitem[Bardeen(1973)]{Bar:1973:BlaHol:}
J.~M. Bardeen, \enquote{Timelike and Null Geodesics in the {K}err Metric,} in
  \emph{{Black Holes}}, edited by C.~D. Witt, and B.~S.~D. Witt, Gordon and
  Breach, New York--London--Paris, 1973, p. 215.

\bibitem[Stuchl{\'{\i}}k(2001)]{Stu:2001:RAGtime2and3:}
Z.~Stuchl{\'{\i}}k, \enquote{Accretion processes in black-hole spacetimes with
  a repulsive cosmological constant,} in \emph{Proceedings of RAGtime 2/3:
  Workshops on black holes and neutron stars, Opava, 11--13/8--10 October
  2000/01}, edited by S.~Hled{\'{\i}}k, and Z.~Stuchl{\'{\i}}k, Silesian
  University in Opava, Opava, 2001, pp. 129--167.

\bibitem[Slan{\'y}(2001)]{Sla:2001:RAGtime2and3:}
P.~Slan{\'y}, \enquote{Some aspects of {K}err--de~{S}itter spacetimes relevant
  to accretion processes,} in \emph{Proceedings of RAGtime 2/3: Workshops on
  black holes and neutron stars, Opava, 11--13/8--10 October 2000/01}, edited
  by S.~Hled{\'{\i}}k, and Z.~Stuchl{\'{\i}}k, Silesian University in Opava,
  Opava, 2001, pp. 119--127.

\bibitem[Misner et~al.(1973)]{Mis-Tho-Whe:1973:Gra:}
C.~W. Misner, K.~S. Thorne, and J.~A. Wheeler, \emph{Gravitation}, Freeman, San
  Francisco, 1973.

\bibitem[Abramowicz et~al.(1988)]{Abr-Car-Las:1988:GENRG2:}
M.~A. Abramowicz, B.~Carter, and J.~Lasota, \emph{Gen. Relativity Gravitation}
  \textbf{20}, 1173 (1988).

\bibitem[Abramowicz et~al.(1995)]{Abr-Nur-Wex:1995:CLAQG:}
M.~A. Abramowicz, P.~Nurowski, and N.~Wex, \emph{Classical Quantum Gravity}
  \textbf{12}, 1467 (1995).

\bibitem[Abramowicz(1990)]{Abr:1990:MONNR:}
M.~A. Abramowicz, \emph{Monthly Notices Roy. Astronom. Soc.} \textbf{245},
  733--746 (1990).

\bibitem[Abramowicz(1992)]{Abr:1992:MONNR:}
M.~A. Abramowicz, \emph{Monthly Notices Roy. Astronom. Soc.} \textbf{256},
  710--718 (1992).

\bibitem[Abramowicz and Bi{\v{c}}{\'a}k(1991)]{Abr-Bic:1991:GENRG2:}
M.~A. Abramowicz, and J.~Bi{\v{c}}{\'a}k, \emph{Gen. Relativity Gravitation}
  \textbf{23}, 941 (1991).

\bibitem[Abramowicz and Miller(1990)]{Abr-Mil:1990:MONNR:}
M.~A. Abramowicz, and J.~C. Miller, \emph{Monthly Notices Roy. Astronom. Soc.}
  \textbf{245}, 729 (1990).

\bibitem[Miller(1993)]{Mil:1993:RenGRCos:}
J.~C. Miller, \enquote{Relativistic Gravitational Collapse,} in \emph{The
  Renaissance of General Relativity and Cosmology}, edited by G.~Ellis,
  A.~Lanza, and J.~Miller, Cambridge University Press, Cambridge, 1993, pp.
  73--85, a Survey to Celebrate the 65th Birthday of Dennis Sciama.

\bibitem[Abramowicz and Prasanna(1990)]{Abr-Pra:1990:MONNR:}
M.~A. Abramowicz, and A.~R. Prasanna, \emph{Monthly Notices Roy. Astronom.
  Soc.} \textbf{245}, 720--728 (1990).

\bibitem[Abramowicz et~al.(1993{\natexlab{a}})]{Abr-Nur-Wex:1993:CLAQG:}
M.~A. Abramowicz, P.~Nurowski, and N.~Wex, \emph{Classical Quantum Gravity}
  \textbf{10}, L183 (1993{\natexlab{a}}).

\bibitem[Abramowicz et~al.(1993{\natexlab{b}})]{Abr-Mil-Stu:1993:PHYSR4:}
M.~A. Abramowicz, J.~Miller, and Z.~Stuchl{\'{\i}}k, \emph{Phys. Rev. D}
  \textbf{47}, 1440--1447 (1993{\natexlab{b}}).

\bibitem[Iyer and Prasanna(1993)]{Iye-Pra:1993:CLAQG:}
S.~Iyer, and A.~R. Prasanna, \emph{Classical Quantum Gravity} \textbf{10},
  L13--L16 (1993).

\bibitem[Stuchl{\'{\i}}k et~al.(2000)]{Stu-Hle-Jur:2000:CLAQG:}
Z.~Stuchl{\'{\i}}k, S.~Hled{\'{\i}}k, and J.~Jur{\'a}{\v{n}}, \emph{Classical
  Quantum Gravity} \textbf{17}, 2691--2718 (2000).

\bibitem[Aguirregabiria et~al.(1996)]{Agu-etal:1996:CLAQG:RaN}
J.~M. Aguirregabiria, A.~Chamorro, K.~R. Nayak, J.~Suinaga, and C.~V.
  Vishveshwara, \emph{Classical Quantum Gravity} \textbf{13}, 2179 (1996).

\bibitem[Stuchl{\'{\i}}k(1990)]{Stu:1990:BULAI:}
Z.~Stuchl{\'{\i}}k, \emph{Bull. Astronom. Inst. Czechoslovakia} \textbf{41},
  341 (1990).

\bibitem[Stuchl{\'{\i}}k and
  Hled{\'{\i}}k(1999{\natexlab{c}})]{Stu-Hle:1999:ACTPS2:}
Z.~Stuchl{\'{\i}}k, and S.~Hled{\'{\i}}k, \emph{Acta Phys. Slovaca}
  \textbf{49}, 795--803 (1999{\natexlab{c}}).

\bibitem[Stuchl{\'{\i}}k and
  Hled{\'{\i}}k(2002{\natexlab{b}})]{Stu-Hle:2002:ACTPS2:}
Z.~Stuchl{\'{\i}}k, and S.~Hled{\'{\i}}k, \emph{Acta Phys. Slovaca}
  \textbf{52}, 363--407 (2002{\natexlab{b}}).

\bibitem[Ernst(1976)]{Ern:1976:JMATP4:}
F.~J. Ernst, \emph{J. Math. Phys.} \textbf{17}, 54 (1976).

\end{thebibliography}

\end{document}